\documentclass[aps,showpacs,twocolumn,nofootinbib]{revtex4}
\usepackage{graphicx}
\usepackage{adjustbox}
\usepackage{dcolumn}
\usepackage{bm}
\usepackage{multirow}
\usepackage{amsmath}
\usepackage[letterpaper, hmargin = {2cm}, vmargin = {2cm}]{geometry}
\usepackage{tablefootnote}

\usepackage{graphicx}
\usepackage{adjustbox}
\usepackage{dcolumn}
\usepackage{bm}
\usepackage{amsmath}
\usepackage{bbold}
\usepackage{multirow}
\usepackage{amsmath}
\usepackage[letterpaper, hmargin = {2cm}, vmargin = {2cm}]{geometry}
\usepackage{tablefootnote}

\usepackage[ulem=normalem]{changes}
\usepackage{mathrsfs}
\usepackage{float}

\usepackage[ulem=normalem]{changes}

\begin{document}

\title{Coriolis coupling effects in  proton-pickup spectroscopic factors from $^{12}$B}

\author{A.~O.~Macchiavelli,  H.~L.~Crawford, R.~M.~Clark, P.~Fallon, I.~Y.~Lee,  C.~Morse, C.~M.~Campbell, M.~Cromaz, and C.~Santamaria}
\affiliation{Nuclear Science Division, Lawrence Berkeley National Laboratory, Berkeley, CA 94720, USA}

\author{J.~Chen}
\affiliation{National Superconducting Cyclotron Laboratory, East Lansing, Michigan 48824, USA}

\author{C.~R.~Hoffman}
\affiliation{Physics Division, Argonne National Laboratory, Argonne, IL 60438, USA}

\author{B.~P.~Kay}
\affiliation{Physics Division, Argonne National Laboratory, Argonne, IL 60438, USA}

\date{\today}
 
\begin{abstract}
Spectroscopic factors to low-lying negative-parity states in $^{11}$Be extracted from the $^{12}$B($d$,$^3$He)$^{11}$Be proton-removal reaction are interpreted within the rotational model.   Earlier predictions of the $p$-wave proton removal strengths in the strong coupling limit of the Nilsson model  underestimated the spectroscopic factors   
to  the  $3/2^-_1$  and $5/2^-_1$ states and suggested that deviations in the $1^+$ ground state of the odd-odd $^{12}$B  due to Coriolis coupling should be further explored.
In this work we use the Particle Rotor Model to take into account these effects and obtain a good description of the level scheme in $^{11}$B, with a moderate $K$-mixing of the proton 
Nilsson levels [110]1/2 and [101]3/2.  This mixing, present in the $1^+$ bandhead of $^{12}$B, is key to explaining the proton pickup data.
\end{abstract}


\maketitle

\section{Introduction}

Shortly after Bohr and Mottelson's development of the collective model~\cite{BM1},  Morinaga showed  that the spectroscopy of a number of nuclei in the $p$-shell could be interpreted 
in terms of rotational bands~\cite{Morinaga56}.    Perhaps one of the most obvious examples is that of $^{8}$Be, as evident from the ground state rotational band and its enhanced $B(E2)$ transition probability~\cite{nndc}.   The strong $\alpha$ clustering in $^{8}$Be naturally suggests that deformation degrees of freedom play a role in the structure of the Be isotopes, a topic that has been extensively discussed in the literature (see Ref.~\cite{VonO} for a review).   

 Turning from a collective model to a Shell Model picture, very nearby to $^{8}$Be, the lightest example of a so-called Island of Inversion~\cite{Poves87,Warburton90} is that at $N$ = 8, where the removal of $p_{3/2}$ protons from $^{14}$C results in a quenching of the $N$=8 shell gap~\cite{Talmi, Sor08, Heyde1, Otsuka20}.   This is evinced in many experimental observations, including the sudden drop of the $E(2^+)$ energy in $^{12}$Be relative to the neighboring even-even isotopes, and the change of the ground state of  $^{11}$Be from the expected 1/2$^-$ to the observed positive parity 1/2$^{+}$  state.  

\begin{figure}[ht!]
\centering\includegraphics[trim=40 0 0 60,clip,width=7cm,angle=90]{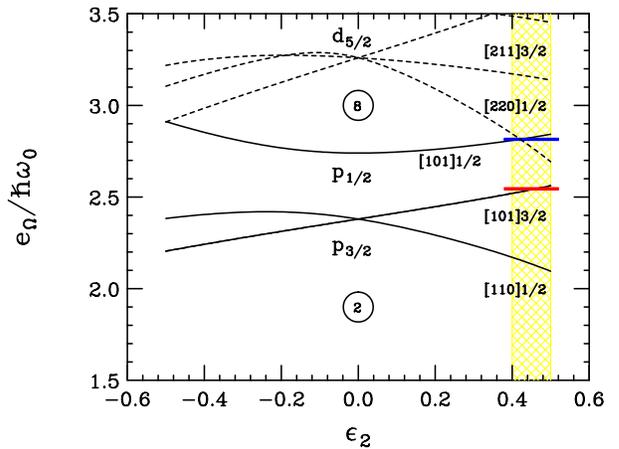}

\caption{(Color online) Nilsson levels relevant for the structure of negative parity neutron states in $^{11}$Be and  $^{11}$B (solid lines). Also shown (dashed lines) are the levels originating from the $d_{5/2}$ spherical orbital. The shaded band indicates the anticipated range of $\epsilon_2$ deformation for these nuclei and the horizontal lines the approximate Fermi levels of the odd neutron (blue) and  the odd  proton (red).  Energies are in units of  the harmonic oscillator frequency, $\hbar\omega_0$. }
\label{fig:fig1}
\end{figure}

 Connecting the collective and single-particle descriptions, Bohr and Mottelson~\cite{BM} actually proposed that the shell inversion could be explained as a result of the convergence of the up-sloping [101]$\tfrac{1}{2}$ and down-sloping [220]$\tfrac{1}{2}$ Nilsson~\cite{Sven, Rag} levels with deformation as seen in Fig.~\ref{fig:fig1}.  Building on  these arguments,  Hamamoto and Shimoura~\cite{Ikuko} explained energy levels and available electromagnetic data on $^{11}$Be and $^{12}$Be in terms of single-particle motion in a deformed potential. It is remarkable that the concept of a deformed  rotating structure appears to be applicable, even when the total number of nucleons is small as in the case of light nuclei.  In fact, level energies and electromagnetic properties that follow the characteristic rotational patterns emerge for $p$-shell nuclei in {\sl ab-initio} no-core configuration interaction calculations~\cite{Mark}.

In a series of articles  we have recently applied  the collective model to understand the structure of nuclei in the  $N$=8 Island of Inversion and spectroscopic factors obtained from direct nucleon addition and removal reactions~\cite{aom1,aom2,aom3}.  The mean-field description seems to capture the main physics ingredients and provides a satisfactory explanation of spectroscopic data, in a simple and intuitive manner.

Here we extend this approach to discuss the results of a recent study of the $^{12}$B($d$,$^3$He)$^{11}$Be reaction~\cite{chen} in terms of the Particle Rotor Model (PRM)~\cite{Rag,Larsson,Semmes}. Estimates of spectroscopic factors in the strong coupling limit, given in Ref.~\cite{chen}, underestimated the experimental data and pointed out that Coriolis effects in the structure of the $1^+$ ground state in $^{12}$B should be taken into account, for which the PRM framework is the one of choice.  We present this analysis here.

\begin{figure}
\centering
\includegraphics[width=6cm]{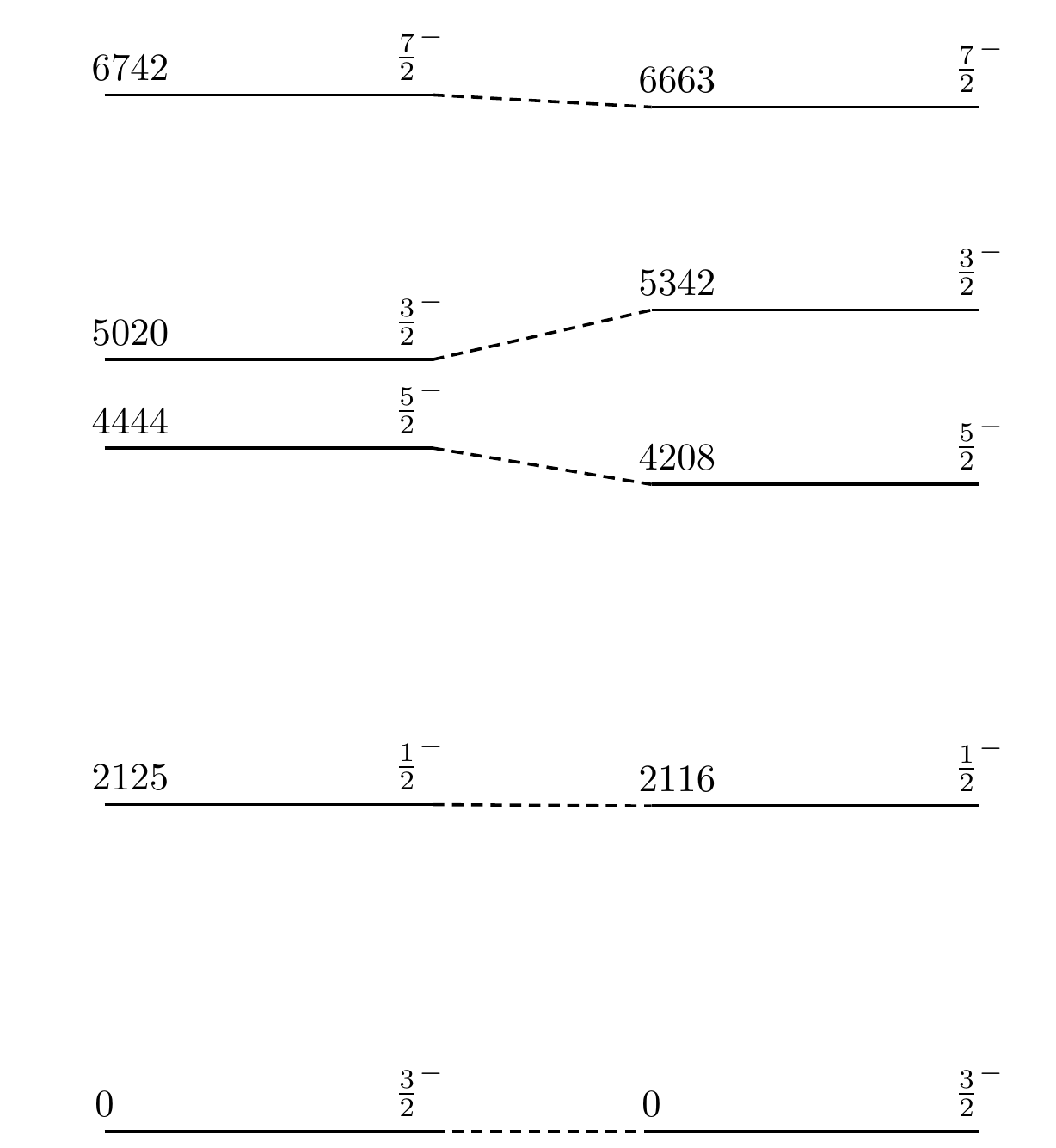}
\caption{ Left: the experimental level scheme of $^{11}$B from Ref.~\cite{nndc}. Right:  Results of the PRM calculations. Energies are in keV.}
\label{fig:fig2}        
\end{figure}

\section{The ground state of $^{12}$B}

In order to assess the structure of the ground state of $^{12}$B, we consider first the odd-neutron and odd-proton  low-lying negative parity states in $^{11}$Be and $^{11}$B respectively.   Considering $^{10}$Be as a core, an inspection of the Nilsson diagram~\cite{Sven} in Fig.~\ref{fig:fig1} suggests that for $N = 7$ the last neutron is expected to occupy the [101]$\tfrac{1}{2}$ level, and for $Z = 5$ the last proton will occupy the [101]$\tfrac{3}{2}$. We have used the standard parameters, $\kappa=0.12$ and  $\mu=0.0$~\cite{Sheline}, and adopt a deformation, $\epsilon_2  \approx $ 0.45 ($\beta_2 \approx 0.6$),  where the crossing of the [220]$\tfrac{1}{2}$ and [101]$\tfrac{1}{2}$ levels is expected to occur, explaining the inversion of the $1/2^{+}$ and the $1/2^{-}$ states in  $^{11}$Be.  

We have previously discussed the positive-parity states in  $^{11}$Be as arising from the strongly coupled [220]$\tfrac{1}{2}$ state~\cite{aom3}, an assignment supported by the calculated gyromagnetic factor $g_K=-2.79$, with decoupling and magnetic-decoupling parameters for the ground state of $a=1.93$ and $b=-1.27$ respectively.  The low-lying negative parity states, namely the $1/2^-_1$, $3/2^-_1$  and $5/2^-_1$  states can be assigned to a  $K=1/2^-$ strongly coupled band built on the neutron [101]$\tfrac{1}{2}$ level, with a decoupling parameter,  $a = 0.5$ in line with the Nilsson predictions.  Further, the  $3/2^-_2$ originates from a neutron hole in the [101]$\tfrac{3}{2}$ level.

The case of $^{11}$B is somewhat different and more complex, requiring an explicit consideration of Coriolis coupling. An attempt to fit the energies of the {\sl yrast} negative parity states,   $3/2^-_1$,  $5/2^-_1$, $7/2^-_1$ ..., to leading order:
\begin{equation} \label{eq:1}
E(I) = E_K + AI(I+1)+ BI^2(I+1)^2+...
\end{equation}
requires an additional term~\cite{BM}  arising from the Coriolis interaction that induces $\Delta K=\pm 2K$ mixing:
\begin{equation} \label{eq:2}
 \Delta E_{rot} = (-1)^{I+K} A_{2K} \frac{(I+K)!}{(I-K)!}, 
\end{equation}  
giving $A=$ 978 keV, $B=$ -17 keV and $A_3 \approx$ 20 keV.
Therefore we carried out a PRM calculation~\cite{Semmes} -- the results, shown in Fig.~\ref{fig:fig2}, are in good agreement with the experimental level scheme.    Here we briefly discuss the physical inputs to the PRM calculation.  We include the three orbits  in Fig.~\ref{fig:fig1} with the Fermi level, $\lambda$, and the pairing gap, $\Delta$, obtained from a BCS calculation using a coupling constant, $G=1.9$ MeV. The solution gives $\lambda=$45.73 MeV and $\Delta=$ 3.3 MeV. The adjusted rotational constant of the core\footnote{In the PRM, we do not include a $BI^2(I+1)^2$ term in the rotational energy but set the $E(2^+)_{core}=A2(2+1)+B(2(2+1))^2$ from Eq.~\ref{eq:1}.} corresponds to a moment of inertia, $ \mathscr{I}$= 0.57 $\hbar^2$/MeV, approximately 60\% of the rigid body value and consistent with the Migdal estimate~\cite{Migdal} for A=11 and the deformation and pairing parameters above. A fit of Eq. (2) to the PRM results gives $A_3 \approx$ 25 keV.\\

The Coriolis $K$-mixing in $^{11}$B gives rise to wave-functions of the form
\begin{equation}
\psi_I  =  \sum_{K}  \mathcal{A}_{IK}  | I K \rangle
\label{eq:eq3}
\end{equation}
in the strong-coupled basis spanned by the intrinsic proton states  [110]$\tfrac{1}{2}$, [101]$\tfrac{3}{2}$, and  [101]$\tfrac{1}{2}$.  The percent contributions (squared amplitudes) of each intrinsic proton state for the states shown in Fig.~\ref{fig:fig2} are given in Table~\ref{tab:tab1}.   The results of the PRM calculations indicate a moderate $K$-mixing for the $3/2_1^-$,  $5/2_1^-$ and $7/2_1^-$, and  the $3/2_2 ^-$ states  with  dominant components of the [101]$\tfrac{3}{2}$  and  [110]$\tfrac{1}{2}$  Nilsson levels in each case.  The  $1/2_1^-$ is essentially a pure [101]$\tfrac{1}{2}$ state.  Note that  the form of the Coriolis matrix elements favors the mixing of the Nilsson levels with $p_{3/2}$ parentage.  It is also worthwhile noting that due to the fact that the Fermi level of the odd proton $^{11}$B is lower than the odd neutron in $^{11}$Be, there is no parity inversion in $^{11}$B and the lowest positive parity state, $1/2^+$, lies at $\approx$ 4.6 MeV relative to the $1/2^-$.

In addition to the reproduction of the energy levels,  the calculated magnetic moment of the ground state is $\mu_{3/2^-}=$ 2.66 $\mu_N$ to be compared to the experimental value of 2.6886489(10) $\mu_N$~\cite{nndc2}.

There is, however, an intriguing discrepancy with the measured $Q_{3/2^-}= 0.04065(26)$eb~\cite{nndc2,11BQ}, which is consistent with the leading order collective model estimate of 0.04 eb, but the PRM result of 0.028 eb
is smaller due to the mixing of the two $I^\pi=3/2^-$ states with $K^\pi=1/2^-$ and $K^\pi=3/2^-$ that have $Q$'s of opposite signs. While one may be tempted to explain this with a larger deformation, $\epsilon_2  \approx $ 0.60~(as in Ref.~\cite{Ikuko}), it would be at the expense of losing the agreement in the energy levels.  This is also the case for the mirror nucleus, $^{11}$C. We do not have an explanation for this discrepancy except to speculate that, perhaps, the deformation is decreasing with spin and the PRM reflects an average. 
The Titled-Axis Cranking model~\cite{Stefan} results for the Be isotopes~\cite{Qi} may support this kind of argument.
It is also interesting to point out that  a recent {\sl ab initio} no-core shell model study~\cite{Petr} of $^{10–14}$B isotopes with realistic $NN$ forces predicts $Q_{3/2^-}$ in the range 0.027 - 0.031 eb (depending on the interaction used), very close to our estimate. \\

\begin{table}
\caption{Coriolis mixing amplitudes in the PRM calculations for $^{11}$B.}
\bigskip
\renewcommand{\arraystretch}{1.2}
\begin{tabular}{c|c|ccc}
\hline\hline
$I^\pi$& Energy&         & $\mathcal{A}_{IK}^2$  [ \% ]         \\
          &     [MeV]                     &    [110]$\tfrac{1}{2}$  & [101]$\tfrac{3}{2}$   & [101]$\tfrac{1}{2}$      \\
 \hline
          $\frac{3}{2}^{-}$& 0.00 &   18             &    80            & 2\\
$\frac{1}{2}^{-}$& 2.12 &      4           &      0          &  96\\
$\frac{5}{2}^{-}$& 4.21       &     6              &    90             & 4\\           
$\frac{3}{2}^{-}$&  5.34       &   81                &       19          &0\\         
 $\frac{7}{2}^{-}$&  6.66     &     45              &      52           & 3\\     
\hline\hline
\end{tabular}
\label{tab:tab1}
\end{table}

In contrast to heavy nuclei, light nuclei are more vulnerable to changes in deformation just by the addition of one particle.  Relevant to our discussions is the fact that $^{12}$C is oblate~\cite{12CQ0} and one question that comes to mind  is why not consider $^{11}B$ as a hole coupled to an oblate core?  In fact, in an early study~\cite{Clegg}, the energy level scheme of $^{11}B$ was described in the collective model with a deformation $\epsilon_2 \approx -0.4$ corresponding to the left side of Nilsson levels shown in Fig.~\ref{fig:fig1}. We have carried out PRM calculations for oblate deformations and obtained an agreement  similar to that in Fig.~\ref{fig:fig2} for $\epsilon_2 \approx -0.35$
and $ \mathscr{I}$= 0.50 $\hbar^2$/MeV. However, this solution gives $Q_{3/2^-} = -0.0063$ eb in clear disagreement with experiment. In looking at the positive parity states, we find that the lowest excitation corresponds to a $5/2^+$ at 
$\approx 10$ MeV from the $3/2^-$ ground state, also inconsistent with the experimental level scheme. \\

Based on the discussion above,  we adopt the prolate results to assess the structure of the ground state 
in $^{12}$B, as a neutron and a proton coupled to the $^{10}$Be core.   The ground state will result from the coupling of the structures  discussed previously, namely a neutron 1/2$^{-}$ based on the [101]$\tfrac{1}{2}$ intrinsic neutron level, and a 3/2$^{-}$ proton state as described in Table~\ref{tab:tab1} , which will give rise to 2$^+$ and 1$^+$ states  with parallel or anti-parallel coupling respectively. 
Following the empirical Gallagher-Moszkowski rule~\cite{GM58}, the lower member of the doublet corresponds to the $1^+$
built from the $K^\pi =0^+, 1^+$ components in agreement with the experimental observations. Further qualitative support comes from the lowest negative parity states, $2^-$ and $1^-$ expected from the coupling to the neutron [220]$\tfrac{1}{2}$. In the oblate side, a $1^-$ to $4^-$ spin multiplet should result from the coupling to the [202]$\tfrac{5}{2}$ level.
Additionally, the measured~\cite{nndc2} magnetic moment, $\mu_{1^-}=$ 1.00306(15) $\mu_N$, and quadrupole moment, $Q_{1^+}=0.0134(14)$ eb, that compare well with leading order estimates,  $\mu_{1^-}=$ 0.97 $\mu_N$, and $Q_{1^+}=0.017$ eb. 
With the ingredients above we proceed to calculate the $p$-wave proton removal strengths, in terms of the Coriolis mixing amplitudes for the $^{12}$B ground state. 

\section{Spectroscopic Factors}

We apply the formalism reviewed in Ref.~\cite{Elbek} to a  proton-pickup reaction, such as $(d,^3$He).  In the strong coupling limit,  the spectroscopic factors ($S_{i,f}$) from an initial ground state $|I_iK_i\rangle$ to a final state $| I_f K_f \rangle$ can be written in terms of the Nilsson amplitudes~\cite{Elbek}:

\begin{equation} 
\begin{split}
\theta_{i, f  }(j\ell,K)& = \langle I_{i}j  K \Omega_\pi  | I_{f} 0\rangle C_{j,\ell} \langle\phi_f|\phi_i\rangle \\
S_{i, f  }& =  \theta_{i, f  }^2(j\ell,K)
\end{split}
\label{eq:eq4}
\end{equation}
where $\langle I_{i}j  K \Omega_\pi  | I_{f} 0\rangle$ is a Clebsch-Gordan coefficient, $C_{j,\ell}$ is the Nilsson wavefunction amplitude,  and $\langle\phi_f|\phi_i\rangle$ is the core overlap between the initial and final states, which we assume to be 1.

Due to the effects of Coriolis coupling discussed  previously   Eq.~\ref{eq:eq4} is generalized to~\cite{Casten71}:
\begin{equation}
S_{i, f} (j\ell) = \big( \sum_{K}  \mathcal{A}_{IK} \theta_{i, f}(j\ell,K) \big)^2
\label{eq:eq5}
\end{equation}

Following from the results in Table~\ref{tab:tab1},  where the 3/2$^-$ ground state is dominated by two contributing Nilsson orbitals, we only consider the [110]$\tfrac{1}{2}$ and [101]$\tfrac{3}{2}$ proton levels which, in the spherical $|j,\ell\rangle$ basis, have the wavefunctions:
\begin{equation}
|[110]\tfrac{1}{2}\rangle=  -0.34 |p_{1/2}\rangle+  0.94|p_{3/2}\rangle 
\label{eq:eq6}
\end{equation}
\begin{equation}
|[101]\tfrac{3}{2}\rangle=   |p_{3/2}\rangle 
\label{eq:eq7}
\end{equation} 

When applied to the case   of $^{12}$B, the PRM hamiltonian for the $1^+$ ground state is a 3x3 matrix in the basis:
\begin{align}
|1\rangle =& |\nu[101]\tfrac{1}{2}\otimes \pi[110]\tfrac{1}{2}\rangle_{K=0} \nonumber  \\ 
|2\rangle =& |\nu[101]\tfrac{1}{2}\otimes \pi[110]\tfrac{1}{2}\rangle_{K=1} \\
| 3\rangle =& |\nu[101]\tfrac{1}{2}\otimes \pi[101]\tfrac{3}{2}\rangle_{K=1} \nonumber
\end{align}
giving a wavefunction of the form:
\begin{align}
|^{12}\textrm{B},1^+ \rangle_\textrm{g.s.} =~&\mathscr{A}_{1}|1\rangle + \mathscr{A}_{2}| 2\rangle   
 + \mathscr{A}_{3}| 3 \rangle 
\label{eq:eq9}
\end{align}

The amplitudes, $\mathscr{A}_{1-3}$  
were fit using  a least-squares minimization to the experimental spectroscopic factor data, yielding $\mathscr{A}_1$ = -0.60(3), $\mathscr{A}_2$ = 0.70(3) and $\mathscr{A}_3$ = 0.40(4) given by the normalization condition $\mathscr{A}_1^2 + \mathscr{A}_2^2$ +$\mathscr{A}_3^2$=1. The derived amplitudes confirm that Coriolis mixing is required to explain the experimental data, reflecting the PRM results for the $3/2^-$ band in $^{11}$B, since the [101]$\tfrac{1}{2}$ neutron can be seen to act as a spectator. The Coriolis effects appear to be somewhat larger in $^{12}$B, which may suggest a core with smaller deformation and a reduced momenta of inertia, both favoring the increased mixing.

We note that within our framework, pickup to the $3/2_2^-$ is not possible since a neutron hole 
in the [101]$\tfrac{3}{2}$ level is not present in the ground state of $^{12}$B. In any case, a contribution to the $3/2_2^- + 5/2_1^-$ doublet due to $K-$mixing in $^{11}$Be is expected to be quite small.

The calculated relative $S_{if}$ in the strong coupling limit and the PRM are compared to the experimental data in Table~\ref{tab:tab2}, which also includes those for the $^{12}$C($p,2p)^{11}$B reaction~\cite{Panin16}.   As already mentioned,  the spectroscopic factors of the $3/2_1^-$ and $5/2_1^-$ states were underestimated in the strong coupling limit and the inclusion of Coriolis coupling appears to solve the discrepancy,  bringing the collective model predictions in line with those of the shell-model and the {\sl ab-initio} Variational Monte Carlo results discussed in Ref.~\cite{chen}. It would be of interest if the study of Ref.~\cite{Petr} could be extended to obtain spectroscopic factors for the reactions in Table~\ref{tab:tab2}.

\begin{table}[ht]
\caption{Comparison between the experimental $\ell=1$ proton removal spectroscopic factors, the Nilsson strong coupling limit and  the PRM results. Note that these are relative to the transitions normalized to 1 and, as such, quenching effects largely cancel.} 
\bigskip
\renewcommand{\arraystretch}{1.4}
\begin{tabular}{c|c|c|ccc}
\hline\hline
Initial & Final & Energy &  &        $S_{i,f}$&\\
State & State & [MeV]  &  Exp  &    Strong Coupling           &Coriolis\footnote{Note that for $^{12}$B,
having two data points and two unknowns, the minimization solution reproduces the data exactly.}\footnote{For $^{12}$C these are based on the $^{11}$B amplitudes in Table.~\ref{tab:tab2} }  \\
\hline 
$^{12}$B & $^{11}$Be&  &\\
1$^{-}_1$ & $\frac{1}{2}^{-}$& 0.32&1 &1&1\\
                          & $\frac{3}{2}^{-}$& 2.35 &2.6(10)&0.8&2.6\\           
                          & $\frac{5}{2}^{-}$  &3.89 &1.7(6) &0.2&1.7\\         
\hline                     
$^{12}$C & $^{11}$B& & & \\
$ 0^+_1$ & $\frac{3}{2}^{-}$ & 0.00&1&1&1\\
                            &  $\frac{1}{2}^{-}$ & 2.12&0.12(2)&0.9&0.1\\
 &  $\frac{3}{2}^{-}$  & 5.02&0.10(2)&0.9&0.3\\
                  \hline\hline
\end{tabular}
\label{tab:tab2}
\end{table}

\section{Conclusion}
We have analyzed spectroscopic factors extracted from the $^{12}$B($d$,$^3$He)$^{11}$Be proton-removal reaction in the framework of the collective model.
The PRM quantitatively explains the available structure data  in $^{11}$Be and $^{11}$B and provides clear evidence of Coriolis coupling in the ground state of $^{12}$B.
The resulting $K$-mixing in the wave-function is key to understand the experimental (relative) spectroscopic factors,  which are underestimated in the strong coupling limit.  
The amplitudes,  empirically adjusted to reproduce the data, are in agreement with the PRM expectations. An application of our phenomenological description to the structure of $^{12}$B (Eqs.~8 and \ref{eq:eq9}) with the two-particle plus rotor model would be an interesting extension to explore.

\begin{acknowledgments}
We would like to thank Profs. Ikuko Hamamoto and Stefan Frauendorf for enlightening discussions on the Nilsson and Particle Rotor Models.  This material is based upon work supported by the U.S. Department of Energy, Office of Science, Office of Nuclear Physics under Contracts No. DE-AC02-05CH11231 (LBNL)
 and DE-AC02-06CH11357 (ANL).  
 \end{acknowledgments}


\end{document}